\documentclass[twocolumn,aps,prd,unsortedaddress,%
superscriptaddress,a4paper,nofootinbib,balancelastpage,preprintnumbers,showpacs,floatfix]{revtex4-1}
\usepackage{latexsym,amsbsy,amsmath}
\usepackage{graphicx}
\usepackage{multirow}
\usepackage[utf8]{inputenc}
\usepackage{soul,xcolor}
\def\vec#1{\boldsymbol{#1}}
\def\ssspin#1#2{\boldsymbol{\sigma}_{#1}.\boldsymbol{\sigma}_{#2}}
\def\llcol#1#2{\tilde{\lambda}_{#1}.\tilde{\lambda}_{#2}}
\def\llss#1#2{\tilde{\lambda}_{#1}.\tilde{\lambda}_{#2}\,\boldsymbol{\sigma}_{#1}\boldsymbol{\sigma}_{#2}}

\begin{document}
\setstcolor{blue}
\title{Pentaquarks with anticharm or beauty revisited}
\author{Jean-Marc~Richard} 
\email{j-m.richard@ipnl.in2p3.fr} 
\affiliation{Universit\'e de Lyon, Institut de Physique Nucl\'eaire de Lyon,
IN2P3-CNRS--UCBL,\\ 
4 rue Enrico Fermi, 69622  Villeurbanne, France} 

\author{A.~Valcarce} 
\email{valcarce@usal.es} 
\affiliation{Departamento de F\'\i sica Fundamental and IUFFyM,\\ 
Universidad de Salamanca, E-37008 Salamanca, Spain}

\author{J.~Vijande} 
\email{javier.vijande@uv.es} 
\affiliation{Unidad Mixta de Investigaci\'on en Radiof{\'\i}sica e Instrumentaci\'on Nuclear en Medicina (IRIMED), \\
Instituto de Investigaci\'on Sanitaria La Fe (IIS-La Fe) \\
Universitat de Valencia (UV) and IFIC (UV-CSIC), Valencia, Spain}

\date{\today} 

\begin{abstract}\noindent
We use a constituent model to analyze the stability of pentaquark $\bar Q qqqq$ configurations with a heavy antiquark $\bar c$ or $\bar b$, and four light quarks
$uuds$, $ddsu$ or $ssud$. The interplay between chromoelectric and chromomagnetic effects is not favorable, and, as a consequence, no bound state is found below the lowest dissociation threshold. 
\end{abstract}

\maketitle 

\section{Introduction}
\label{se:intro}
There is a renewed interest in the spectroscopy of exotic hadrons containing one or two heavy constituents. For a review, see, e.g.,~\cite{Chen:2016qju,Richard:2016eis,Ali:2017jda,Albuquerque:2016znh}. New configurations are studied, and exotic states suggested in the 70s or 80s are revisited. 

Among the first multiquark candidates involving heavy flavors, there is the $\bar Q qqqq$ pentaquark proposed independently and simultaneously by the Grenoble group and by Harry Lipkin~\cite{Gignoux:1987cn,Lipkin:1987sk}.
The word {\em pentaquark} was introduced in this context.

For the chromomagnetic interaction, the $P_{\bar Q}=\bar Q qqqq$ is very similar to the $H=uuddss$ of Jaffe~\cite{Jaffe:1976yi}, who realized that for a spin $J=0$ and color-singlet state, the color-spin operator 
\begin{equation} \label{eq:cm1}
 \mathcal{O}_n=\sum_{i<j}^n \llss{i}{j}~,
\end{equation}
applied to $uuddss$, reaches its largest eigenvalue $\mathcal{O}_6=24$, to be compared to $\mathcal{O}_3=8$ for each  spin 1/2 baryon of the threshold. This means that in the limit of flavor symmetry SU(3)$_F$, a chromomagnetic operator $H_{CM}=-a\,\mathcal{O}_6$ gives an additional downwards shift 
\begin{equation} \label{eq:cm2}
 \delta M_{CM}=8 \,a=\frac12\,(\Delta-N)\sim 150\,\mathrm{MeV}~,
\end{equation}
as compared to the threshold, provided the short-range correlation factor (in simple potential models, it is proportional to the expectation value of $\delta(\vec r_{ij})$) is assumed to be the \emph{same} for the $H$ as for the ground-state baryons. 

Similarly, an eigenvalue $\mathcal{O}_4=16$ is found for a $qqqq$ system in a state of color $3$ and spin $J_q=0$ corresponding to a SU(3)$_F$ triplet of flavor. This means that in the limit where the mass of the heavy quark becomes infinite, i.e., the chromomagnetic energy is restricted to the light sector, a downwards shift $\delta M=8\, a\sim 150\,$MeV is obtained for $\bar Qqqqq$, as compared to its lowest threshold $\bar Qq+qqq$. Again, the value $\delta M\sim 150\,$MeV is derived assuming that the $qq$ short-range correlation is the same in $\bar Qqqqq$ as in $qqq$.

This mechanism of chromomagnetic binding was analyzed in several subsequent papers~\cite{Karl:1987uf,Fleck:1989ff,Leandri:1989su,Yuan:2012wz}.\footnote{Sometimes, e.g., in~\cite{Fleck:1989ff}, the ordering of the $\bar D\Lambda$ vs.~$\bar D_s p$ threshold was not discussed}\@  When the SU(3)$_F$ symmetry is broken, the pentaquark is penalized (say for fixed mass $m$ of  $u$ and $d$, and increased mass $m_s$ for the strange quark). Adopting a finite mass for
the heavy quark  also goes against the stability of the heavy pentaquark~\cite{Fleck:1989ff,Leandri:1989su}.

Note that the reasoning leading to $\delta M=8\,a$ for a spin 1/2  $\bar Qqqqq$ predicts a chromomagnetic binding $\delta M=16\,a/3\sim 100\,$MeV for the spin 3/2 state. Hence both spin $s=1/2$ and $s=3/2$ states deserve some investigation. 

In this letter we adopt a generic constituent model, containing chromoelectric and chromomagnetic contributions, tuned to reproduce the masses of the mesons and baryons entering the various thresholds and study the pentaquark configurations $\bar Q uuds$, $\bar Q ddsu$ and $\bar Q ssdu$ with $Q=c$ or $b$, for both $s=1/2$ and $s=3/2$, using a powerful variational method.  We switch on and off some of the contributions to understand why stability is hardly reached. 

The paper is organized as follows. In Sec.~\ref{se:mod}, we present briefly the model and the variational method. The results are shown in Sec.~\ref{se:resu}. Some further comments are proposed in Sec.~\ref{se:conc}
\section{Model}\label{se:mod}
We adopt the so-called AL1 model by Semay and Silvestre-Brac \cite{Semay:1994ht}, already used in a number of exploratory studies of multiquark systems, for instance in our recent investigation of the hidden-charm sector $\bar c c qqq$~\cite{Richard:2017una} or doubly-heavy tetraquarks $QQ\bar q\bar q$~\cite{Richard:2018yrm}. It includes a standard Coulomb-plus-linear central potential, supplemented by a smeared version of the chromomagnetic interaction,
\begin{align}\label{eq:mod1}
V(r)  &=  -\frac{3}{16}\, \llcol{i}{j}
\left[\lambda\, r - \frac{\kappa}{r}-\Lambda + \frac{V_{SS}(r)}{m_i \, m_j}  \, \ssspin{i}{j}\right] \, ,\\
V_{SS}  &= \frac{2 \, \pi\, \kappa^\prime}{3\,\pi^{3/2}\, r_0^3} \,\exp\left(- \frac{r^2}{r_0^2}\right) ~,\quad
 r_0 =  A \left(\frac{2 m_i m_j}{m_i+m_j}\right)^{-B}\!,\nonumber
 \end{align}
where
$\lambda=$ 0.1653 GeV$^2$, $\Lambda=$ 0.8321 GeV, $\kappa=$ 0.5069, $\kappa^\prime=$ 1.8609,
$A=$ 1.6553 GeV$^{B-1}$, $B=$ 0.2204, $m_u=m_d=$ 0.315 GeV, $m_s=$ 0.577 GeV, $m_c=$ 1.836 GeV and $m_b=$ 5.227 GeV. 
Here, $\llcol{i}{j}$ is a color factor, suitably modified for the quark-antiquark pairs.
We disregard the small three-body term of this model used in~\cite{Semay:1994ht} to fine-tune the baryon masses vs.\ the meson masses.
Note that the smearing parameter of the spin-spin term is adapted to the masses involved in the quark-quark or quark-antiquark pairs. 
It is worth to emphasize that the parameters of the AL1 potential are constrained in a simultaneous fit of
36 well-established meson states and 53 baryons,
with a remarkable agreement with data, as could be seen in Table 2 of Ref.~\cite{Semay:1994ht}.

Before implementing any constraint of symmetry, a system $\bar q_1 q_2q_3q_4q_5$ has three possible color components for an overall color singlet, five spin components for a spin $s=1/2$, and four for $s=3/2$. The configurations with $s=5/2$ do not support any bound state in the simple chromomagnetic model and thus 
are not further studied in the present paper. As for color, a singlet $\bar q_1 q_2$ is associated with a $q_3q_4q_5$ singlet, and a $\bar q_1 q_2$ octet can be neutralized by any of the two $q_3q_4q_5$ octets. Three alternative bases can be obtained by replacing $q_2$ by either $q_3$, $q_4$ or~$q_5$. 

In the limit of large $m_Q$, attention was focused in the configuration $q_2q_3q_4q_5$ with optimal chromomagnetic attraction. It corresponds to a color 3 and spin $s_{2345}=0$, which is a combination of the state with $s_{23}=s_{45}=0$ and the one with $s_{23}=s_{45}=1$. For finite $m_Q$, the three spin states with $s_{2345}=1$ also contribute, that can match $s=1/2$ when coupled to $s_1=1/2$. These three latter states with $s_{2345}=1$ allow one to build an overall $s=3/2$, as well as the quark state $s_{2345}=2$. 

We calculate the binding energy of mesons, baryons and pentaquarks by means of an expansion on a set of correlated Gaussians, schematically
\begin{equation}\label{eq:mod2}
 \Psi_\alpha(\vec x_1, \ldots)=\sum_{i=1}^N \gamma_i \left[\exp(-\tilde{X}.A_i.X/2)\pm \cdots\right]~\, ,
\end{equation}
where the ellipses stand for terms deduced by permutations dictated by the symmetries of the system. The subscript $\alpha$ refers to the spin-isospin-color  components which are coupled by the interaction \eqref{eq:mod1}. The vector $X$ stands for the set of Jacobi coordinates describing the relative motion, namely $\tilde{X}=\{x_1, \ldots, \vec x_{n-1}\}$ for a $n$-body system. The matrices $A_i$ are symmetric and definite positive. The weight factors $\gamma_i$ and the range matrices $A_i$ are tuned by standard techniques to minimize the energy, for an increasing number of terms $N$, until a reasonable convergence is reached. 
We push our calculation until the difference of introducing a new term is smaller than 2 MeV.

In principle, the results are independent of the choice of any particular set of the Jacobi coordinates
for the five-quark problem shown in Fig. 1 of Ref.~\cite{Richard:2017una}. However, some sets lead to 
matrices $A_i$ which are closer to a diagonal form and thus leads to faster convergence to the
lowest eigenvalue. Thus, changing the set of Jacobi coordinates and the initial values of the parameters 
entering the matrices $A_i$ is a routine consistency check of such variational methods that has
been carried in the present study as well as in Ref.~\cite{Richard:2017una}.

The first concern is whether or not a state is bound below the lowest threshold, say $MB$, where $M$ is a meson, and $B$ a baryon. An immediate strategy is to detect the ground state energy lower than the threshold energy $M+B$. If the state is  unbound, one observes a slow decrease toward $M+B$ as $N$ increases. It turns out useful to look also at the content of the variational wave  function, which comes very close to 100\% in the singlet-singlet channel of color in the $MB$ basis. On the other hand, if a variational state converges to a bound state as $N$ increases, then it includes sizable hidden-color components even for low $N$.
\section{Results}\label{se:resu}
\subsection{Results for \boldmath $\bar Quuds$\unboldmath}
A calculation of the masses of mesons $D(c\bar u)$, \dots, $B_s(s\bar b)$ and baryons $p(uud)$, \dots $\Lambda_b(bud)$ leads to the threshold masses shown in Table~\ref{tab:thr1}, which also displays the best 5-body energy with the required convergence, $N=5$ or $N=6$ in Eq.~\eqref{eq:mod2}.
\begin{table}[ht]
\caption{\label{tab:thr1}Threshold masses for $\bar c uuds$ and $\bar b uuds$, and pentaquark theoretical estimate, $5q$. All masses are in GeV.}
 \begin{ruledtabular}
  \begin{tabular}{ccccc}
         & $\bar D\,\Lambda$ & 3.016 & $B\,\Lambda$ & 6.447\\
 $J=1/2$ & $\bar D_s\,p$ & 2.958 & $B_s\,p$ & 6.357 \\
         &  $5q$             & 2.966 &  $5q$      & 6.361 \\
\hline  
         & $\bar D^*\,\Lambda$ & 3.170 & $B^*\,\Lambda$ & 6.504\\
 $J=3/2$ & $\bar D^*_s\,p$ & 3.098 & $B_s^*\,p$ & 6.413 \\
         &  $5q$             & 3.101 &  $5q$      & 6.418 \\
  \end{tabular}
 \end{ruledtabular}
\end{table}
No binding is found, as seen from the variational energy remaining above the threshold and from the color-content of the variational wave function,
100\% in the lowest $MB$ channel of the threshold. This negative result survives a number of changes in the model, by modifying some parameters.

One of these checks consists in recalculating the threshold and pentaquark energies with the strength of the hyperfine interaction, the parameter $\kappa'$ in Eq.~\eqref{eq:cm1}, artificially increased by a factor $f'$, i.e., $\kappa'\to f'\,\kappa'$. As expected from the 1987 papers \cite{Gignoux:1987cn,Lipkin:1987sk}, stability should be reached for large $f'$, when the chromomagnetic interaction dominates. Stability is reached for $f'\sim 3$ in the bottom case and
$f'\sim 1.8$ in the charm sector, as expected from the $1/(m_qM_Q)$ dependence of the chromomagnetic interaction. This means that the short-range correlation is significantly weakened in the pentaquark as compared to its value in baryons, and that the configuration favoring the chromomagnetic binding is not optimal when the chromoelectric and kinetic terms are included. 

As a further check, one can also play with the central potential. For instance, if the Coulomb term in \eqref{eq:mod1} is reduced by a factor of 10, namely $\kappa\to \kappa/10$, and the  threshold and pentaquark energies are recalculated, then the pentaquark remains unbound, but the critical factor for forcing binding is slightly reduced, to $f'\sim 2.6$ in the bottom sector. As already noticed in \cite{Fleck:1989ff}, this  indicates that there is somewhat a conflict between the chromoelectric and the chromomagnetic contributions: the chromoelectric forces favor some internal configuration that is nearly orthogonal to the one optimizing the chromomagnetic term. 
\subsection{Results for \boldmath$\bar Q ssud$\unboldmath}
The calculations described above are now repeated for the pentaquark with strangeness $S=-2$. The results are shown in Table \ref{tab:thr2}.
The same tests as for $S=-1$ have been carried out, which confirm the absence of binding for this kind of modeling, independently from the details of the tuning of the parameters. 
The factor $f'$ which ensures binding when multiplying the chromomagnetic term, is now about 2.6 in the bottom sector, i.e., somewhat smaller than in the case of strangeness $S=-1$~\cite{Leandri:1989su}, but still indicating that the pentaquark with one heavy antiquark and strangeness $S=-2$ is rather far from stability in this class of potential models. 
\begin{table}[!th]
\caption{\label{tab:thr2}Threshold masses for $\bar c ssud$ and $\bar b ssud$, and pentaquark theoretical estimate, $5q$. All masses are in GeV.
}
 \begin{ruledtabular}
  \begin{tabular}{ccccc}
         & $\bar D_s\,\Lambda$ & 3.116 & $B_s\,\Lambda$ & 6.515\\
 $J=1/2$ & $\bar D\,\Xi$ & 3.243 & $B\,\Xi$ & 6.674 \\
         &  $5q$             & 3.174 &  $5q$      & 6.567 \\
\end{tabular}
 \end{ruledtabular}
\end{table}
\subsection{Other flavor configurations}
Some calculations were carried out for the case of hidden strangeness, namely $\bar s uuds$. In this case, the ordering of the thresholds are inverted as compared to the values observed in Table~\ref{tab:thr1}. For $Q=s$, $K\Lambda$ is below $\eta p$, while for $Q=c$, $\bar D \Lambda$ is above $\bar D_s p$. No bound states were found.

As a curiosity, we have run the case of a fictitious charm quark, say $c'$,  of mass 1\,GeV, i.e., intermediate between the actual charm quark and the strange quark. No fine tuning is done to fix that mass. It is now observed that the thresholds $\bar c' u+uds$ and $\bar c' s+uud$ are nearly degenerate. This provides, in principle, the opportunity to gain some attraction by mixing the configurations corresponding to each threshold. However, one does not get binding, because the coupling between the two configurations $\bar{c}^{\;\prime} u+uds$ and $\bar{c}^{\;\prime} s+uud$ is not strong enough.

\section{Outlook}\label{se:conc}
Various configurations have been studied for the heavy pentaquark systems $\bar Qqqqq$ with $Q=c$ or $b$ and $qqqq=uuds$, $ddsu$ and $ssud$, and spin $s=1/2$ or $s=3/2$, in the framework of a conventional constituent model. No bound state has been obtained, nor any indication for some narrow resonance in the continuum.
Our results are on the line of the recent experimental findings of the LHCb collaboration~\cite{Aaij:2017jgf}.

To perform exploratory studies of systems with more than 
three-quarks it is of basic importance to work with models that correctly
describe the two- and three-quark problems which thresholds are made 
of. Therefore, varying the parameters 
do not significantly affect our results, as we have checked, because 
the induced changes in the multiquark and threshold energies are similar.

Obviously, a multiquark state
contains color configurations that are not present asymptotically in the thresholds
and this is the basic ingredient that may drive to a bound state. As already 
emphasized in Ref.~\cite{Richard:2017una}, and shown in Figs. 2 or 3 of this 
reference, there is a strong competition between the color-spin configurations favored 
by the chromoelectric terms and the ones favored by the chromomagnetic terms, 
and this mismatch spoils the possible 
binding of pentaquarks with anticharm or beauty.

We have explored different possibilities for the mass $m_Q$ of the heavy quark. 
Increasing the mass of the heavy quark, separates the two thresholds as 
seen in Table~\ref{tab:thr1} when
comparing the results for charm and bottom cases. A large $m_Q$ induces a 
large chromoelectric attraction in the $\bar Q s$ pair, but the same 
attraction is present in the lowest threshold, $B_s p$. On the contrary, 
a mass $m_Q\simeq 1\,$GeV makes the two thresholds, $B \Lambda$ and
$B_s p$, nearly degenerate. Then, one may expect 
a favorable mixing. However, there is a conflict between 
the color-spin configurations favored by the chromoelectric terms and the 
chromomagnetic ones.

For the resonances, our conclusion is based on the content of the variational wavefunctions, 
which are found to consist mainly of two color-singlets in the channel corresponding to the lowest threshold. 

For the light pentaquark states and for the  hidden-flavor sector $\bar Q Q qqq$ the method of 
real scaling has been used, which is rather demanding in terms of 
computation \cite{Hiyama:2005cf,Yamaguchi:2017zmn,Hiyama:2018ukv}. Certainly, a critical comparison of the  different methods of handling resonances is in order.  

\acknowledgments
This work has been funded by Ministerio de Econom\'\i a, Industria y Competitividad
and EU FEDER under Contract No.\ FPA2016-77177.

%

\begin{thebibliography}{17}%
\makeatletter
\providecommand \@ifxundefined [1]{%
 \@ifx{#1\undefined}
}%
\providecommand \@ifnum [1]{%
 \ifnum #1\expandafter \@firstoftwo
 \else \expandafter \@secondoftwo
 \fi
}%
\providecommand \@ifx [1]{%
 \ifx #1\expandafter \@firstoftwo
 \else \expandafter \@secondoftwo
 \fi
}%
\providecommand \natexlab [1]{#1}%
\providecommand \enquote  [1]{``#1''}%
\providecommand \bibnamefont  [1]{#1}%
\providecommand \bibfnamefont [1]{#1}%
\providecommand \citenamefont [1]{#1}%
\providecommand \href@noop [0]{\@secondoftwo}%
\providecommand \href [0]{\begingroup \@sanitize@url \@href}%
\providecommand \@href[1]{\@@startlink{#1}\@@href}%
\providecommand \@@href[1]{\endgroup#1\@@endlink}%
\providecommand \@sanitize@url [0]{\catcode `\\12\catcode `\$12\catcode
  `\&12\catcode `\#12\catcode `\^12\catcode `\_12\catcode `\%12\relax}%
\providecommand \@@startlink[1]{}%
\providecommand \@@endlink[0]{}%
\providecommand \url  [0]{\begingroup\@sanitize@url \@url }%
\providecommand \@url [1]{\endgroup\@href {#1}{\urlprefix }}%
\providecommand \urlprefix  [0]{URL }%
\providecommand \Eprint [0]{\href }%
\providecommand \doibase [0]{http://dx.doi.org/}%
\providecommand \selectlanguage [0]{\@gobble}%
\providecommand \bibinfo  [0]{\@secondoftwo}%
\providecommand \bibfield  [0]{\@secondoftwo}%
\providecommand \translation [1]{[#1]}%
\providecommand \BibitemOpen [0]{}%
\providecommand \bibitemStop [0]{}%
\providecommand \bibitemNoStop [0]{.\EOS\space}%
\providecommand \EOS [0]{\spacefactor3000\relax}%
\providecommand \BibitemShut  [1]{\csname bibitem#1\endcsname}%
\let\auto@bib@innerbib\@empty
\bibitem [{\citenamefont {Chen}\ \emph {et~al.}(2016)\citenamefont {Chen},
  \citenamefont {Chen}, \citenamefont {Liu},\ and\ \citenamefont
  {Zhu}}]{Chen:2016qju}%
  \BibitemOpen
  \bibfield  {author} {\bibinfo {author} {\bibfnamefont {H.-X.}\ \bibnamefont
  {Chen}}, \bibinfo {author} {\bibfnamefont {W.}~\bibnamefont {Chen}}, \bibinfo
  {author} {\bibfnamefont {X.}~\bibnamefont {Liu}}, \ and\ \bibinfo {author}
  {\bibfnamefont {S.-L.}\ \bibnamefont {Zhu}},\ }\href {\doibase
  10.1016/j.physrep.2016.05.004} {\bibfield  {journal} {\bibinfo  {journal}
  {Phys. Rept.}\ }\textbf {\bibinfo {volume} {639}},\ \bibinfo {pages} {1}
  (\bibinfo {year} {2016})},\ \Eprint {http://arxiv.org/abs/1601.02092}
  {arXiv:1601.02092 [hep-ph]} \BibitemShut {NoStop}%
\bibitem [{\citenamefont {Richard}(2016)}]{Richard:2016eis}%
  \BibitemOpen
  \bibfield  {author} {\bibinfo {author} {\bibfnamefont {J.-M.}\ \bibnamefont
  {Richard}},\ }\href {\doibase 10.1007/s00601-016-1159-0} {\bibfield
  {journal} {\bibinfo  {journal} {Few Body Syst.}\ }\textbf {\bibinfo {volume}
  {57}},\ \bibinfo {pages} {1185} (\bibinfo {year} {2016})},\ \bibinfo {note}
  {{Special issue for the 30th anniversary of Few-Body Systems}},\ \Eprint
  {http://arxiv.org/abs/1606.08593} {arXiv:1606.08593 [hep-ph]} \BibitemShut
  {NoStop}%
\bibitem [{\citenamefont {Ali}\ \emph {et~al.}(2017)\citenamefont {Ali},
  \citenamefont {Lange},\ and\ \citenamefont {Stone}}]{Ali:2017jda}%
  \BibitemOpen
  \bibfield  {author} {\bibinfo {author} {\bibfnamefont {A.}~\bibnamefont
  {Ali}}, \bibinfo {author} {\bibfnamefont {J.~S.}\ \bibnamefont {Lange}}, \
  and\ \bibinfo {author} {\bibfnamefont {S.}~\bibnamefont {Stone}},\ }\href
  {\doibase 10.1016/j.ppnp.2017.08.003} {\bibfield  {journal} {\bibinfo
  {journal} {Prog. Part. Nucl. Phys.}\ }\textbf {\bibinfo {volume} {97}},\
  \bibinfo {pages} {123} (\bibinfo {year} {2017})},\ \Eprint
  {http://arxiv.org/abs/1706.00610} {arXiv:1706.00610 [hep-ph]} \BibitemShut
  {NoStop}%
\bibitem [{\citenamefont {Albuquerque}\ \emph {et~al.}(2016)\citenamefont
  {Albuquerque}, \citenamefont {Narison}, \citenamefont {Fanomezana},
  \citenamefont {Rabemananjara}, \citenamefont {Rabetiarivony},\ and\
  \citenamefont {Randriamanatrika}}]{Albuquerque:2016znh}%
  \BibitemOpen
  \bibfield  {author} {\bibinfo {author} {\bibfnamefont {R.}~\bibnamefont
  {Albuquerque}}, \bibinfo {author} {\bibfnamefont {S.}~\bibnamefont
  {Narison}}, \bibinfo {author} {\bibfnamefont {F.}~\bibnamefont {Fanomezana}},
  \bibinfo {author} {\bibfnamefont {A.}~\bibnamefont {Rabemananjara}}, \bibinfo
  {author} {\bibfnamefont {D.}~\bibnamefont {Rabetiarivony}}, \ and\ \bibinfo
  {author} {\bibfnamefont {G.}~\bibnamefont {Randriamanatrika}},\ }\href
  {\doibase 10.1142/S0217751X16501967} {\bibfield  {journal} {\bibinfo
  {journal} {Int. J. Mod. Phys.}\ }\textbf {\bibinfo {volume} {A31}},\ \bibinfo
  {pages} {1650196} (\bibinfo {year} {2016})},\ \Eprint
  {http://arxiv.org/abs/1609.03351} {arXiv:1609.03351 [hep-ph]} \BibitemShut
  {NoStop}%
\bibitem [{\citenamefont {Gignoux}\ \emph {et~al.}(1987)\citenamefont
  {Gignoux}, \citenamefont {Silvestre-Brac},\ and\ \citenamefont
  {Richard}}]{Gignoux:1987cn}%
  \BibitemOpen
  \bibfield  {author} {\bibinfo {author} {\bibfnamefont {C.}~\bibnamefont
  {Gignoux}}, \bibinfo {author} {\bibfnamefont {B.}~\bibnamefont
  {Silvestre-Brac}}, \ and\ \bibinfo {author} {\bibfnamefont {J.~M.}\
  \bibnamefont {Richard}},\ }\href {\doibase 10.1016/0370-2693(87)91244-5}
  {\bibfield  {journal} {\bibinfo  {journal} {Phys. Lett.}\ }\textbf {\bibinfo
  {volume} {B193}},\ \bibinfo {pages} {323} (\bibinfo {year}
  {1987})}\BibitemShut {NoStop}%
\bibitem [{\citenamefont {Lipkin}(1987)}]{Lipkin:1987sk}%
  \BibitemOpen
  \bibfield  {author} {\bibinfo {author} {\bibfnamefont {H.~J.}\ \bibnamefont
  {Lipkin}},\ }\href {\doibase 10.1016/0370-2693(87)90055-4} {\bibfield
  {journal} {\bibinfo  {journal} {Phys. Lett.}\ }\textbf {\bibinfo {volume}
  {B195}},\ \bibinfo {pages} {484} (\bibinfo {year} {1987})}\BibitemShut
  {NoStop}%
\bibitem [{\citenamefont {Jaffe}(1977)}]{Jaffe:1976yi}%
  \BibitemOpen
  \bibfield  {author} {\bibinfo {author} {\bibfnamefont {R.~L.}\ \bibnamefont
  {Jaffe}},\ }\href {\doibase 10.1103/PhysRevLett.38.195} {\bibfield  {journal}
  {\bibinfo  {journal} {Phys. Rev. Lett.}\ }\textbf {\bibinfo {volume} {38}},\
  \bibinfo {pages} {195} (\bibinfo {year} {1977})},\ \bibinfo {note} {[Erratum:
  Phys. Rev. Lett.38,617(1977)]}\BibitemShut {NoStop}%
\bibitem [{\citenamefont {Karl}\ and\ \citenamefont
  {Zenczykowski}(1987)}]{Karl:1987uf}%
  \BibitemOpen
  \bibfield  {author} {\bibinfo {author} {\bibfnamefont {G.}~\bibnamefont
  {Karl}}\ and\ \bibinfo {author} {\bibfnamefont {P.}~\bibnamefont
  {Zenczykowski}},\ }\href {\doibase 10.1103/PhysRevD.36.3520} {\bibfield
  {journal} {\bibinfo  {journal} {Phys. Rev.}\ }\textbf {\bibinfo {volume}
  {D36}},\ \bibinfo {pages} {3520} (\bibinfo {year} {1987})}\BibitemShut
  {NoStop}%
\bibitem [{\citenamefont {Fleck}\ \emph {et~al.}(1989)\citenamefont {Fleck},
  \citenamefont {Gignoux}, \citenamefont {Richard},\ and\ \citenamefont
  {Silvestre-Brac}}]{Fleck:1989ff}%
  \BibitemOpen
  \bibfield  {author} {\bibinfo {author} {\bibfnamefont {S.}~\bibnamefont
  {Fleck}}, \bibinfo {author} {\bibfnamefont {C.}~\bibnamefont {Gignoux}},
  \bibinfo {author} {\bibfnamefont {J.~M.}\ \bibnamefont {Richard}}, \ and\
  \bibinfo {author} {\bibfnamefont {B.}~\bibnamefont {Silvestre-Brac}},\ }\href
  {\doibase 10.1016/0370-2693(89)90797-1} {\bibfield  {journal} {\bibinfo
  {journal} {Phys. Lett.}\ }\textbf {\bibinfo {volume} {B220}},\ \bibinfo
  {pages} {616} (\bibinfo {year} {1989})}\BibitemShut {NoStop}%
\bibitem [{\citenamefont {Leandri}\ and\ \citenamefont
  {Silvestre-Brac}(1989)}]{Leandri:1989su}%
  \BibitemOpen
  \bibfield  {author} {\bibinfo {author} {\bibfnamefont {J.}~\bibnamefont
  {Leandri}}\ and\ \bibinfo {author} {\bibfnamefont {B.}~\bibnamefont
  {Silvestre-Brac}},\ }\href {\doibase 10.1103/PhysRevD.40.2340} {\bibfield
  {journal} {\bibinfo  {journal} {Phys. Rev.}\ }\textbf {\bibinfo {volume}
  {D40}},\ \bibinfo {pages} {2340} (\bibinfo {year} {1989})}\BibitemShut
  {NoStop}%
\bibitem [{\citenamefont {Yuan}\ \emph {et~al.}(2012)\citenamefont {Yuan},
  \citenamefont {Wei}, \citenamefont {He}, \citenamefont {Xu},\ and\
  \citenamefont {Zou}}]{Yuan:2012wz}%
  \BibitemOpen
  \bibfield  {author} {\bibinfo {author} {\bibfnamefont {S.~G.}\ \bibnamefont
  {Yuan}}, \bibinfo {author} {\bibfnamefont {K.~W.}\ \bibnamefont {Wei}},
  \bibinfo {author} {\bibfnamefont {J.}~\bibnamefont {He}}, \bibinfo {author}
  {\bibfnamefont {H.~S.}\ \bibnamefont {Xu}}, \ and\ \bibinfo {author}
  {\bibfnamefont {B.~S.}\ \bibnamefont {Zou}},\ }\href {\doibase
  10.1140/epja/i2012-12061-2} {\bibfield  {journal} {\bibinfo  {journal} {Eur.
  Phys. J.}\ }\textbf {\bibinfo {volume} {A48}},\ \bibinfo {pages} {61}
  (\bibinfo {year} {2012})},\ \Eprint {http://arxiv.org/abs/1201.0807}
  {arXiv:1201.0807 [nucl-th]} \BibitemShut {NoStop}%
\bibitem [{\citenamefont {Semay}\ and\ \citenamefont
  {Silvestre-Brac}(1994)}]{Semay:1994ht}%
  \BibitemOpen
  \bibfield  {author} {\bibinfo {author} {\bibfnamefont {C.}~\bibnamefont
  {Semay}}\ and\ \bibinfo {author} {\bibfnamefont {B.}~\bibnamefont
  {Silvestre-Brac}},\ }\href {\doibase 10.1007/BF01413104} {\bibfield
  {journal} {\bibinfo  {journal} {Z. Phys.}\ }\textbf {\bibinfo {volume}
  {C61}},\ \bibinfo {pages} {271} (\bibinfo {year} {1994})}\BibitemShut
  {NoStop}%
\bibitem [{\citenamefont {Richard}\ \emph {et~al.}(2017)\citenamefont
  {Richard}, \citenamefont {Valcarce},\ and\ \citenamefont
  {Vijande}}]{Richard:2017una}%
  \BibitemOpen
  \bibfield  {author} {\bibinfo {author} {\bibfnamefont {J.~M.}\ \bibnamefont
  {Richard}}, \bibinfo {author} {\bibfnamefont {A.}~\bibnamefont {Valcarce}}, \
  and\ \bibinfo {author} {\bibfnamefont {J.}~\bibnamefont {Vijande}},\ }\href
  {\doibase 10.1016/j.physletb.2017.10.036} {\bibfield  {journal} {\bibinfo
  {journal} {Phys. Lett.}\ }\textbf {\bibinfo {volume} {B774}},\ \bibinfo
  {pages} {710} (\bibinfo {year} {2017})},\ \Eprint
  {http://arxiv.org/abs/1710.08239} {arXiv:1710.08239 [hep-ph]} \BibitemShut
  {NoStop}%
\bibitem [{\citenamefont {Richard}\ \emph {et~al.}(2018)\citenamefont
  {Richard}, \citenamefont {Valcarce},\ and\ \citenamefont
  {Vijande}}]{Richard:2018yrm}%
  \BibitemOpen
  \bibfield  {author} {\bibinfo {author} {\bibfnamefont {J.-M.}\ \bibnamefont
  {Richard}}, \bibinfo {author} {\bibfnamefont {A.}~\bibnamefont {Valcarce}}, \
  and\ \bibinfo {author} {\bibfnamefont {J.}~\bibnamefont {Vijande}},\ }\href
  {\doibase 10.1103/PhysRevC.97.035211} {\bibfield  {journal} {\bibinfo
  {journal} {Phys. Rev.}\ }\textbf {\bibinfo {volume} {C97}},\ \bibinfo {pages}
  {035211} (\bibinfo {year} {2018})},\ \Eprint
  {http://arxiv.org/abs/1803.06155} {arXiv:1803.06155 [hep-ph]} \BibitemShut
  {NoStop}%
\bibitem [{\citenamefont {Aaij}\ \emph {et~al.}(2018)\citenamefont {Aaij} \emph
  {et~al.}}]{Aaij:2017jgf}%
  \BibitemOpen
  \bibfield  {author} {\bibinfo {author} {\bibfnamefont {R.}~\bibnamefont
  {Aaij}} \emph {et~al.} (\bibinfo {collaboration} {LHCb}),\ }\href {\doibase
  10.1103/PhysRevD.97.032010} {\bibfield  {journal} {\bibinfo  {journal} {Phys.
  Rev.}\ }\textbf {\bibinfo {volume} {D97}},\ \bibinfo {pages} {032010}
  (\bibinfo {year} {2018})},\ \Eprint {http://arxiv.org/abs/1712.08086}
  {arXiv:1712.08086 [hep-ex]} \BibitemShut {NoStop}%
\bibitem [{\citenamefont {Hiyama}\ \emph {et~al.}(2006)\citenamefont {Hiyama},
  \citenamefont {Kamimura}, \citenamefont {Hosaka}, \citenamefont {Toki},\ and\
  \citenamefont {Yahiro}}]{Hiyama:2005cf}%
  \BibitemOpen
  \bibfield  {author} {\bibinfo {author} {\bibfnamefont {E.}~\bibnamefont
  {Hiyama}}, \bibinfo {author} {\bibfnamefont {M.}~\bibnamefont {Kamimura}},
  \bibinfo {author} {\bibfnamefont {A.}~\bibnamefont {Hosaka}}, \bibinfo
  {author} {\bibfnamefont {H.}~\bibnamefont {Toki}}, \ and\ \bibinfo {author}
  {\bibfnamefont {M.}~\bibnamefont {Yahiro}},\ }\href {\doibase
  10.1016/j.physletb.2005.11.086} {\bibfield  {journal} {\bibinfo  {journal}
  {Phys. Lett.}\ }\textbf {\bibinfo {volume} {B633}},\ \bibinfo {pages} {237}
  (\bibinfo {year} {2006})},\ \Eprint {http://arxiv.org/abs/hep-ph/0507105}
  {arXiv:hep-ph/0507105 [hep-ph]} \BibitemShut {NoStop}%
\bibitem{Yamaguchi:2017zmn}
Y. Yamaguchi, A. Giachino, A. Hosaka, E. Santopinto,
  S. Takeuchi, and M. Takizawa.
\newblock Phys. Rev. {\bf D96}, 114031 (2017).
\bibitem [{\citenamefont {Hiyama}\ \emph {et~al.}(2018)\citenamefont {Hiyama},
  \citenamefont {Hosaka}, \citenamefont {Oka},\ and\ \citenamefont
  {Richard}}]{Hiyama:2018ukv}%
  \BibitemOpen
  \bibfield  {author} {\bibinfo {author} {\bibfnamefont {E.}~\bibnamefont
  {Hiyama}}, \bibinfo {author} {\bibfnamefont {A.}~\bibnamefont {Hosaka}},
  \bibinfo {author} {\bibfnamefont {M.}~\bibnamefont {Oka}}, \ and\ \bibinfo
  {author} {\bibfnamefont {J.-M.}\ \bibnamefont {Richard}},\ }\href@noop {} {\
  {\bibinfo
  {journal} {Phys. Rev.}\ }\textbf {\bibinfo {volume} {C98}},\ \bibinfo {pages}
  {045208} (\bibinfo {year} {2018})},\ \Eprint
  {http://arxiv.org/abs/1803.11369}
  {arXiv:1803.11369 [nucl-th]} \BibitemShut {NoStop}%
\end{thebibliography}
%

\end{document}